\documentclass{article}
\usepackage{tabularray}
\usepackage{graphicx}
\usepackage{amsmath}
\usepackage{color}
\usepackage{algorithm}
\usepackage{algpseudocode}
\usepackage{algorithmicx}
\usepackage{amssymb}
\definecolor{Alto}{rgb}{0.858,0.858,0.858}
\usepackage{xcolor}
\usepackage[dvipsnames]{xcolor}
\newcommand{\figref}[1]{Fig.~{\color{red}{\ref{#1}}}}

\usepackage{graphicx}
\usepackage{subcaption}
\usepackage{bbm}
\usepackage{amsmath} 
\usepackage{booktabs}  
\usepackage{multirow} 
\usepackage{xspace}

\definecolor{turquoise}{cmyk}{0.65,0,0.1,0.3}
\definecolor{purple}{rgb}{0.65,0,0.65}
\definecolor{dark_green}{rgb}{0, 0.5, 0}
\definecolor{orange}{rgb}{0.8, 0.6, 0.2}
\definecolor{red}{rgb}{0.8, 0.2, 0.2}
\definecolor{darkred}{rgb}{0.6, 0.1, 0.05}
\definecolor{blueish}{rgb}{0.0, 0.3, .6}
\definecolor{blue}{rgb}{0, 0.3, 1}
\definecolor{light_gray}{rgb}{0.7, 0.7, .7}
\definecolor{pink}{rgb}{1, 0, 1}
\definecolor{greyblue}{rgb}{0.25, 0.25, 1}
\definecolor{light_gray}{gray}{0.95}
\definecolor{light-green}{rgb}{0.82, 0.94, 0.75}

\newcommand{\paragrapht}[1]{\noindent\textbf{#1}}  % tidy \paragraph

\newcommand{\ie}{\textit{i.e.}}

\newcommand{\method}{\textsc{ACVIS}\xspace}

\usepackage{tikz}

\usepackage{spconf,amsmath,graphicx,hyperref}
\usepackage{amsfonts}
\usepackage{color}
\usepackage{tabularray}
\UseTblrLibrary{booktabs}
\usepackage[section]{placeins}
\usepackage{color}
\usepackage{hyperref}

\usepackage{xcolor}
\hypersetup{
  colorlinks=true,
  linkcolor=black,
  citecolor=blue!85!black,
}

\title{Learning What to hear: Audio-centric Queries and Counting for AVIS}
\title{Learning What to hear: Boosting Sound-Source Association \\ for Robust Audiovisual Instance Segmentation}

\name{Jinbae Seo\(^{1}\), Hyeongjun Kwon\(^{1}\), Kwonyoung Kim\(^{1}\), Jiyoung Lee\(^{2*}\), and Kwanghoon Sohn\(^{1,3*}\)\thanks{* Corresponding authors.}}
\address{$^1$Yonsei University \quad $^2$School of AI and Software, Ewha Womans University \\ $^3$Korea Institute of Science and Technology (KIST)}
\begin{document}
%\ninept
%
\maketitle
\begin{abstract}
Audiovisual instance segmentation (AVIS) aims to accurately localize and track sounding objects throughout video sequences.
Existing methods suffer from visual bias stemming from two fundamental issues: uniform additive fusion prevents queries from specializing to different sound sources, while visual-only training objectives limit queries from converging to arbitrary salient objects.
We propose Audio-Centric Query Generation (ACQG) using cross-attention mechanism, enabling each query to selectively attend to distinct sound sources and carry sound-specific priors into visual decoding.
Additionally, we introduce Sound-Aware Ordinal Counting (SAOC) loss that explicitly supervises sounding object numbers through ordinal regression with monotonic consistency constraints, preventing visual-only convergence during training.
Experiments on AVISeg benchmark demonstrate consistent improvements: +1.64 mAP, +0.6 HOTA, and +2.06 FSLA, validating that query specialization and explicit counting supervision are crucial for accurate audiovisual instance segmentation.
Our code and models are available at \url{https://github.com/jinbae-s/ACVIS}
\end{abstract}

\vspace{-0.5em}
\section{Introduction}
\label{sec:intro}

\begin{figure}[t]
    \centering
    \includegraphics[width=\linewidth]{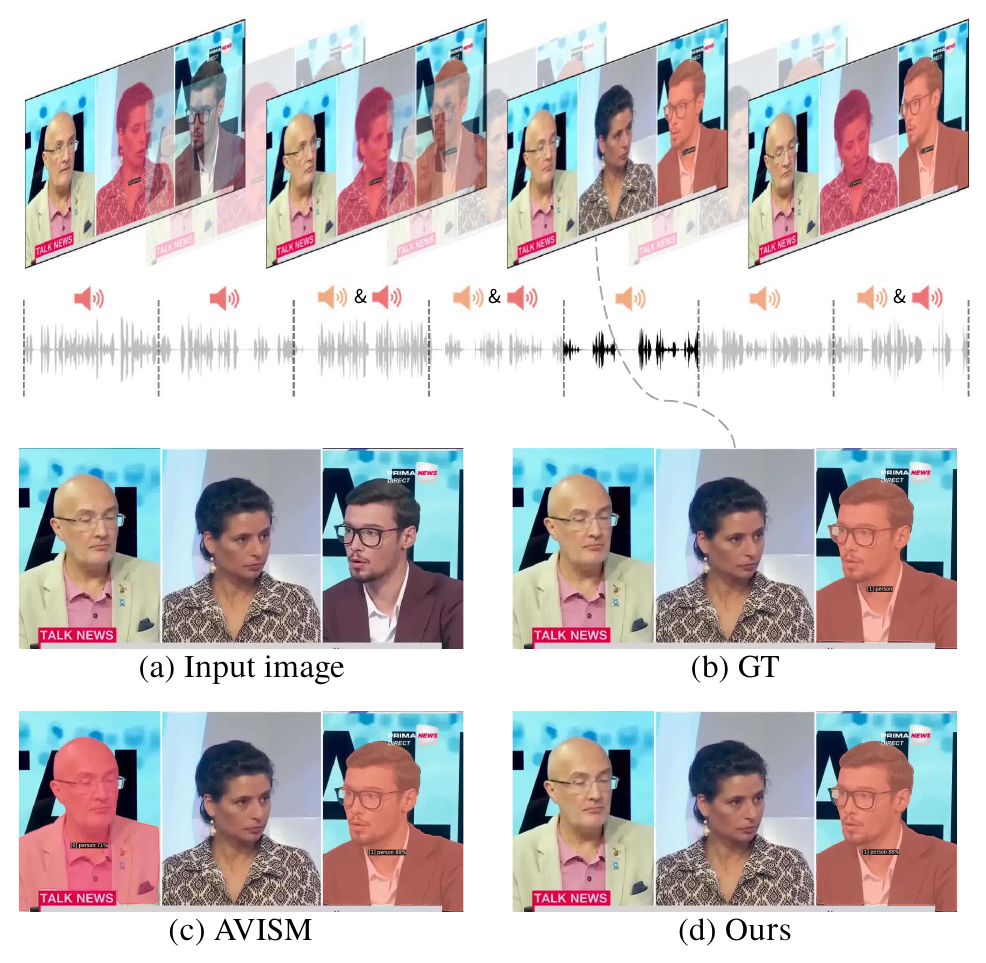}
    \vspace{-2.5em}
    \caption{Visual bias in audiovisual instance segmentation. While ground truth (b) indicates only one person speaking, previous work (AVISM)~\cite{guo2025avis} (c) detects two visible people due to visual dominance. 
    Our \method (d) correctly identifies the speaking person by maintaining audio-visual balance through specialized queries and counting supervision.}
    \label{fig:1}
    \vspace{-1.5em}
\end{figure}

\vspace{-0.5em}
Humans effortlessly perceive complex scenes by integrating what they see with what they hear.
In many real-world scenarios, such as identifying a speaking person in a crowded video, or distinguishing between overlapping instruments in a performance, sound provides critical cues that vision alone cannot resolve as shown in \figref{fig:1}.
This motivates the task of audiovisual instance segmentation (AVIS), which aims to segment object instances in the visual scene with their associated audio sources.
% Recognizing and segmenting visual scenes via audio signals represents a fundamental pursuit in audiovisual scene understanding.

Early audiovisual learning explored multimodal correspondence through self-supervision~\cite{arandjelovic2017look}, evolving from coarse localization~\cite{senocak2018learning} to attention-based~\cite{qian2020multiple} and contrastive methods~\cite{chen2021localizing}.
%,owens2018audio ,arandjelovic2018objects,mo2022localizing
However, these methods are limited to producing only heat maps or bounding boxes rather than semantic-level masks.
To address this limitation, audiovisual segmentation~\cite{zhou2022avs,zhou2025avss} has emerged with pixel-level precision.
Subsequent works—AVSegFormer~\cite{gao2024avsegformer} with transformers, SAMA-AVS~\cite{liu2024annotation} leverages SAM~\cite{kirillov2023segany}, and VCT~\cite{huang2025vctavs} with vision-centric queries—have improved semantic segmentation but all remained limited to semantic-level, unable to distinguish individual instances.
% audiovisual segmentation methods emerged with pixel-level precision.
% AVS~\cite{zhou2022avs,zhou2025avss} has pioneered pixel-level segmentation with the temporal pixel-wise audiovisual interaction module, but only at the semantic level, unable to distinguish between individual instances of the same category.
% Subsequent works have improved semantic segmentation through various approaches: AVSegFormer~\cite{gao2024avsegformer} adapted transformer architectures for better audiovisual fusion, SAMA-AVS~\cite{liu2024annotation} leveraged the segment anything model (SAM)~\cite{kirillov2023segany} with lightweight adapter, and VCT~\cite{huang2025vctavs} introduced a vision-centric paradigm that uses visually-derived queries to overcome the perception ambiguity caused by mixed audio signals.
% Despite these advances in semantic segmentation, instance-level discrimination has remained unsolved.
Recently, AVISM~\cite{guo2025avis} has achieved instance-level segmentation through a compacted two-stage architecture: frame-level object localization followed by video-level object tracking.
In the first stage, the model predicts frame queries corresponding to instances for each frame by adding audio features to learnable query tokens, which are then refined through visual cross-attention to produce audiovisual frame queries.
In the second stage, these frame-level detections are associated across time through a tracker that establishes temporal correspondences between instances.
While this approach enables instance-level tracking, the frame-level localizer suffers from a critical limitation: audio features are uniformly integrated into all queries through a simple addition process, preventing queries from specializing to different sound sources.

We address the aforementioned limitations through two complementary innovations.
First, we replace additive fusion with cross-attention to enable each query to selectively attend to different sound sources in the audio signal.
To facilitate better audiovisual correspondence in subsequent decoder layers, our method produces specialized audio-centric queries where each query is pre-assigned to specific audio patterns.
Although this modification promotes capturing sound-related instance queries, matching sound sources to instances in complex real-world scenarios, where multiple instances with the same semantic label are loud simultaneously, is highly challenging. 
Therefore, we introduce a sound-aware ordinal counting (SAOC) loss that provides the missing audio-centric constraint.
Explicitly supervising how many queries should activate for sounding objects ensures the decoder optimization considers both visual appearance and audio presence, preventing convergence to visual-only solutions.
Our contributions are summarized as:
\vspace{-0.7em}
\begin{itemize}
    \item We introduce a novel \textbf{A}udio-\textbf{C}entric audio\textbf{V}isual \textbf{I}nstance \textbf{S}egmentation (\method) for sound source-aware AVIS, introducing an audio-centric query generator (ACQG) with sound-aware ordinal counting (SAOC) loss.
    \vspace{-0.7em}
    \item Our \method highlights the robustness of query discrimination according to the sound source by ordinal regression through guidance of counting audible objects.
    \vspace{-0.7em}
    % specialization to distinct sound sources via cross-attention, facilitating visual correspondence matching for audio events.
    % \item We propose Sound-Aware Ordinal Counting loss that explicitly supervises sounding object numbers through ordinal regression, preventing visual-only convergence during the object localizer optimization.
    \item We demonstrate through experiments where improvements of +1.64 mAP, +0.6 HOTA, and +2.06 FSLA, validating our frame-level innovations as crucial for accurate AVIS.
    \vspace{-0.7em}
\end{itemize}

\vspace{-0.5em}
\section{Method}
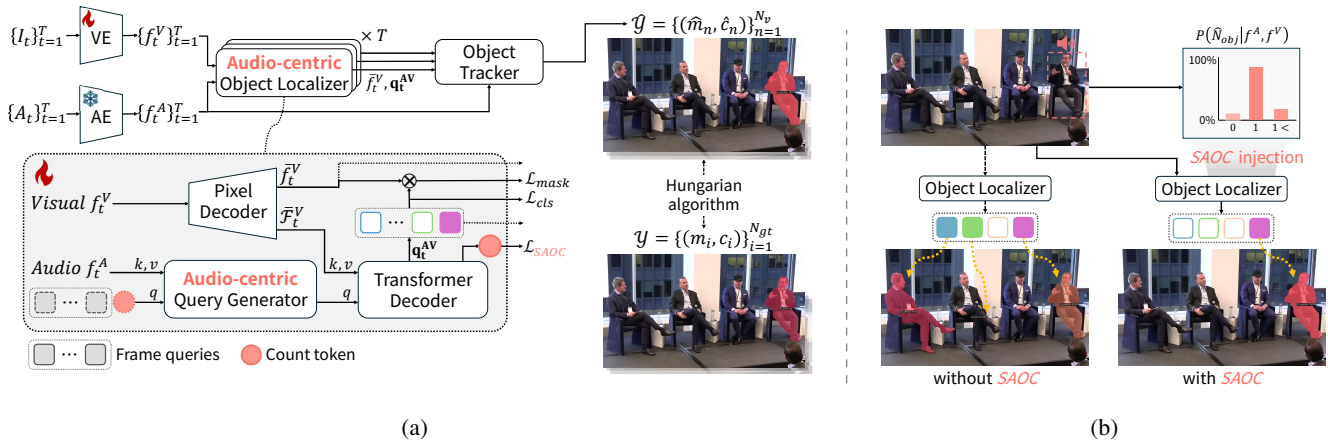
\begin{figure*}[!ht]
    \centering
    \begin{subfigure}[b]{0.62\linewidth}
        \centering
        \includegraphics[width=\linewidth]{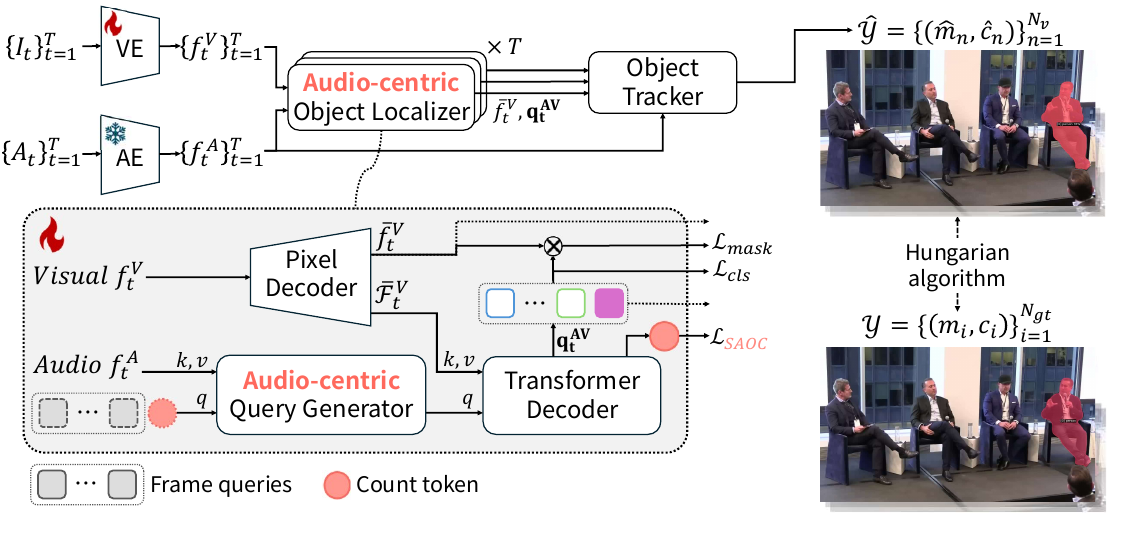}
        % \vspace{-1.2em}
        \vspace{-0.5em}
        \caption{}
        \label{fig:2a}
    \end{subfigure}
    \hfill
    \begin{tikzpicture}[remember picture, overlay]
        \draw[dashed, gray, line width=0.5pt] (0,0.8) -- (0,5.5);
    \end{tikzpicture}
    \hfill
    \begin{subfigure}[b]{0.34\linewidth}
        \centering
        \includegraphics[width=\linewidth]{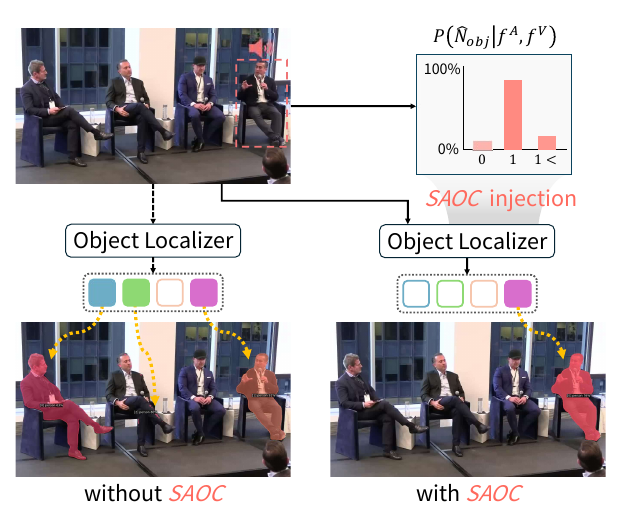}
        \vspace{-1.6em}
        \caption{}
        \label{fig:2b}
    \end{subfigure}
    \vspace{-1.2em}
     \caption{(a) Overall architecture with audio-centric object localizer and object tracker. Frame queries and count token are processed through our query generator and decoder for AVIS. (b) Our SAOC loss prevents visual bias: without our loss (left), the model over-detects visually salient objects; with our loss (right), only sounding objects are segmented.}
    \vspace{-1.2em}
    \label{fig:overall}
\end{figure*}

\vspace{-0.4em}
\subsection{Problem Formulation and Overview}
\vspace{-0.1em}
AVIS aims to classify, segment, and track all sounding objects in a given video.
Given an input video with audio, the model predicts a set of instance masks with associated class labels.

% Our method takes AVISM~\cite{guo2025avis} as a baseline model, which follows the set-prediction paradigm~\cite{carion2020detr,cheng2022masked} with a two-stage architecture: frame-level object localizer and video-level object tracker.
Our method takes AVISM~\cite{guo2025avis} as a baseline, which follows the set-prediction paradigm~\cite{carion2020detr,cheng2022masked} with a two-stage architecture: frame-level object localizer and video-level object tracker.
% Formally, audio and video encoders extract audio feature $f^A_t$ and visual feature $f^V_t$ for a given $t$-th segment input, respectively.
Formally, audio and video encoders extract audio feature $f^A_t$ and visual feature $f^V_t$ for a given $t$-th segment, respectively.
The pixel decoder within the object localizer generates enhanced multi-scale features: $\bar{f}^V_t$ for final-resolution map and $\bar{\mathcal{F}}^V_t$ for multi-scale representations. 
These features, combined with audio features $f^A_t$ and learnable queries $\mathbf{q}$, produce audiovisual frame queries $\mathbf{q}^{AV}_t$ at frame $t$.
The object tracker aggregates these frame queries $\{\mathbf{q}^{AV}_t\}_{t=1}^T$ into $N_v$ video queries to generate final predictions $\hat{\mathcal{Y}}$:
\begin{align}
   &\mathbf{q}^{AV}_t = \texttt{ObjLocalizer}(f^V_t,f^A_t,\mathbf{q})   \\
   &\hat{\mathcal{Y}} = \texttt{ObjTracker}(\{\bar{f}^V_t\}^T_{t=1},\
    \{f^A_t\}^T_{t=1},\{\mathbf{q}^{AV}_t\}^T_{t=1}).
\end{align}

However, as exemplified in \figref{fig:1} (c), the baseline often fails to separate sound sources at the instance level.
The uniform additive fusion,
\begin{equation}
\mathbf{q}_{t}^A = \mathbf{q} + \mathbf{1}_{N_f} \otimes f^{A}_{t},\quad \text{where}\ \mathbf{q}\in \ \mathbb{R}^{N_{f}\times D},\ f^{A}_{t}\in\mathbb{R}^{D}
\end{equation}
forces all queries to share identical audio representation, preventing discrimination between instances (e.g., multiple speakers).
Furthermore, we speculate that visually concentrated constraints (\ie, mask and classification losses) do not guarantee query specialization to different sound sources.
We tackle these problems by introducing (i) an audio-centric query generator (ACQG) that conditions learnable queries directly on audio representations and (ii) a sound-aware ordinal counting (SAOC) loss that explicitly guides the model to detect sounding objects rather than arbitrary visual objects.

\vspace{-0.7em}
\subsection{Audio-Centric Frame Queries}
\vspace{-0.1em}
To obtain fine-grained audiovisual correspondence at the frame-level, our \method takes each frame and corresponding audio segment as inputs.
We also set \(N_f\) learnable frame queries \(\mathbf{q}_t\in\mathbb{R}^{N_f\times D}\), which derive instance-wise mask in the segmentation decoder. 
At each time step \(t\), our audio-centric query generator (ACQG) fuses $\mathbf{q}_t$ with the audio feature \(f^{A}_t\in\mathbb{R}^{D}\) to obtain audio-centric frame queries. 
ACQG consists of three cross-attention layers:
\begin{equation}
\mathbf{q}^{A}_t = \texttt{ACQG}(\mathbf{q}_t, f^A_t, f^A_t) \in \mathbb{R}^{N_f \times D},
\end{equation}
where $\mathbf{q}_t$ serves as query and $f^A_t$ as both key and value.
This module enables each query to selectively attend to different patterns in the audio signal.
Each query thereby selectively attends to different sound sources, creating audio-specialized queries that carry sound-specific priors into the audiovisual frame query generation process.

The segmentation decoder~\cite{guo2025avis} then processes audio-centric frame queries with multi-scale visual features:
\begin{equation}
\mathbf{q}^{AV}_t \;=\;  \texttt{Decoder}(\mathbf{q}^{A}_t,\, \mathcal{F}^V_t, \mathcal{F}^V_t) \;\in\; \mathbb{R}^{N_f\times D}\,.
\end{equation}
After processing all frames, the object tracker produces temporally aligned audiovisual frame queries.

\vspace{-0.7em}
\subsection{Sound-Aware Ordinal Counting loss}
\vspace{-0.1em}
To enhance the sound-source awareness in our framework, we design a count token $q_{\text{cnt}} \in \mathbb{R}^{D}$ to optimize the weights with our SAOC loss. 
% The count token is concatenated with frame queries, and aggregates the information about the number of sound sources in the object localizer.
The count token is a learnable embedding that is concatenated with frame queries, and aggregates the information about the number of sound sources in the object localizer.
We denote by $q^{AV}_{cnt}$ the count token after the segmentation decoder.
Following previous work~\cite{shi2023corn}, we model counting sound instances as an ordinal regression problem using conditional probabilities to ensure rank consistency.
% The count token $q_{\text{cnt}}^{AV}$ is processed by $\phi_{\text{cnt}}: \mathbb{R}^{D} \rightarrow \mathbb{R}^{K_{\max}}$ to predict:
% {\color{blue}
The count token $q_{\text{cnt}}^{AV}$ is processed by linear projection head $\phi_{\text{cnt}}: \mathbb{R}^{D} \rightarrow \mathbb{R}^{K_{\max}}$ to predict:
% }
\begin{equation}
\{p_k\}_{k=0}^{K_{\max}-1} = \sigma(\phi_{\text{cnt}}(q_{\text{cnt}}^{AV})),
\end{equation}
where $p_0 = P(\hat{N}_{obj} > 0)$ is the marginal probability and $p_k = P(\hat{N}_{obj} > k | \hat{N}_{obj} > k-1)$ for $k \in \{1, ..., K_{\max}-1\}$ are conditional probabilities.

Given ground-truth count $N_{obj}$, we define ordinal targets $t_k = \mathbbm{1}[N_{obj} > k]$ and compute our SAOC loss:
\begin{equation}
\mathcal{L}_{\text{SAOC}} = -\frac{1}{T}\sum_{t=1}^{T}\sum_{k=0}^{K_{\max}-1} \left[ t_k \log p_k + (1-t_k) \log(1-p_k) \right].
\end{equation}
This ordinal formulation enforces monotonic consistency through conditional structure, ensuring $P(\hat{N}_{obj} > k) \geq P(\hat{N}_{obj} > k+1)$ while providing stable gradients.
By explicitly supervising sounding object counts, SAOC prevents the decoder from activating queries for arbitrary visual objects.

\vspace{-0.7em}
\subsection{Training and Inference}
\vspace{-0.1em}
\paragrapht{Training.} 
The object tracker aggregates frame queries $\{\mathbf{q}^{AV}_t\}_{t=1}^T$ and uses $N_v$ video queries to generate final mask predictions $\hat{\mathcal{Y}}$.
During training, these $N_v$ predictions are matched with ground-truth instance-wise masks via Hungarian algorithm~\cite{kuhn1955hungarian}.
Frame-level auxiliary heads provide direct supervision on the object localizer outputs.
Our \method is optimized with four loss terms, including frame-level masking, video-level masking, frame-video query alignment, and our proposed SAOC losses.
Following AVISM~\cite{guo2025avis}, $\mathcal{L}_{\text{frame}}$ and $\mathcal{L}_{\text{video}}$ supervise frame and video-level predictions via bipartite matching respectively, while $\mathcal{L}_{\text{sim}}$ aligns query embeddings between frame and video-level across temporal scales.
These visual-centric losses alone often cause over-segmentation of salient objects.
To solve this problem, our proposed counting loss $\mathcal{L}_{\text{SAOC}}$ optimizes the network to learn audio-aware visual features with audio-centric supervision.
To sum up, the overall training objective is:
\begin{equation}
\mathcal{L} = \mathcal{L}_{\text{AVIS}} + \lambda_{\text{SAOC}}\mathcal{L}_{\text{SAOC}},
\end{equation}
where $\lambda_\text{SAOC}$ is the hyperparameter for $\mathcal{L}_\text{SAOC}$, and $\mathcal{L}_{\text{AVIS}}$ is the weighted sum of $\mathcal{L}_{\text{frame}}$, $\mathcal{L}_{\text{video}}$ and $\mathcal{L}_{\text{sim}}$, as defined in AVISM~\cite{guo2025avis}.

\paragrapht{Inference.}
% During inference, the $N_v$ video-level predictions are filtered through confidence thresholding to produce $N_{\text{pred}}$ final instance trajectories $\hat{\mathcal{Y}}=\{(\hat{m}_n, \hat{c}_n)\}_{n=1}^{N_{\text{pred}}}$, where each instance spans multiple frames with mask $\hat{m}_n \in [0,1]^{T \times H \times W}$ and class logits $\hat{c}_n \in \mathbb{R}^{K+1}$.
During inference, the $N_v$ video-level predictions are filtered through confidence thresholding to produce $N_{\text{pred}}$ final instance trajectories $\hat{\mathcal{Y}}=\{(\hat{m}_n, \hat{c}_n)\}_{n=1}^{N_{\text{pred}}}$, where each instance spans multiple frames with mask $\hat{m}_n \in [0,1]^{T \times H \times W}$ and class logits $\hat{c}_n \in \mathbb{R}^{N_c+1}$, where $N_c$ is the number of object classes.
\vspace{-0.5em}
\section{Experiments}
\begin{figure*}[t]
\begin{minipage}{0.02\textwidth}
\centering
\rotatebox{90}{\small }
\end{minipage}
\begin{minipage}{0.97\textwidth}
\includegraphics[width=\textwidth]{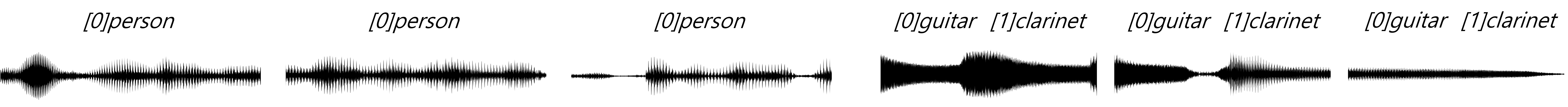}
\end{minipage}\\[0.3em]
\begin{minipage}{0.02\textwidth}
\centering
\rotatebox{90}{\small (a) AVISM}
\end{minipage}
\begin{minipage}{0.97\textwidth}
\includegraphics[width=\textwidth]{ICASSP2026_Paper_Templates/Figure/fig3_avism.pdf}
\end{minipage}\\[0.2em]
\begin{minipage}{0.02\textwidth}
\centering
\rotatebox{90}{\small (b) Ours}
\end{minipage}
\begin{minipage}{0.97\textwidth}
\includegraphics[width=\textwidth]{ICASSP2026_Paper_Templates/Figure/fig3_ours.pdf}
\end{minipage}
\vspace{-.5em}
\caption{Qualitative results across diverse audio scenarios with varying sound sources.}
\vspace{-.5em}
\label{fig:qualitative}
\end{figure*}
\begin{table*}[t]
\centering
% First row: Main results and Backbone comparison
\begin{minipage}[t]{0.45\textwidth}
\centering
% \begin{table}[t]
\centering
\setlength{\tabcolsep}{4pt}
\small{
\begin{tabular}{l ccc ccc}
    \toprule
    Method &  mAP & HOTA & FSLA & FSLAn & FSLAs & FSLAm \\
    \midrule
    AVISM~\cite{guo2025avis} & 45.04 & 64.52 & 44.42 & 20.62 & 32.62 & 54.99 \\
    \method  & 46.68 & 65.12 & 46.48 & 10.74 & 34.45 & 58.81 \\
    \bottomrule
\end{tabular}}
\caption{Performance comparisons on AVISeg.}
\label{tab:main}
% \end{table}

\end{minipage}
\hfill
\begin{minipage}[t]{0.48\textwidth}
\centering
% \begin{table}[t]
\centering
\small{
\begin{tabular}{ll ccc}
    \toprule
    Backbone & Pre-trained dataset & mAP & HOTA & FSLA \\
    \midrule
    ResNet-50    & IN      & 42.14 & 62.09 & 42.87 \\
    ResNet-50    & IN+COCO & 46.68 & 65.12 & 46.48 \\
    Swin-L      & IN+COCO & 54.16 & 72.96 & 54.17     \\
    \bottomrule
\end{tabular}}
\caption{Impact of visual backbone and pre-training dataset.}
\label{tab:backbone}
% \end{table}

\end{minipage}

\vspace{1em}

% Second row: Three ablation studies
\begin{minipage}[t]{0.32\textwidth}
\centering
\centering
\small{
\begin{tabular}{cc ccc}
    \toprule
     ACQG & $\mathcal{L}_{\text{SAOC}}$ & mAP   & HOTA  & FSLA \\
     \midrule
    \checkmark     &            & 45.17 & 63.27 & 45.45 \\
                   & \checkmark & 45.13 & 64.98 & 45.30 \\
    \checkmark     & \checkmark & 46.68 & 65.12 & 46.48 \\
    \bottomrule
\end{tabular}}
\caption{Impact of audio-centric query generator and $\mathcal{L}_{\text{SAOC}}$.}
\vspace{-0.3em}
\label{tab:acqg}
\end{minipage}
\hspace{0.04\textwidth}
\begin{minipage}[t]{0.30\textwidth}
\centering
\centering
\small
\begin{tabular}{c ccc}
    \toprule
    Loss type                   & mAP   & HOTA  & FSLA \\
    \midrule
    $\mathcal{L}_{\text{CE}}$   & 44.45 & 63.95 & 44.00 \\
    $\mathcal{L}_{\text{SAOC}}$ & 46.68 & 65.12 & 46.48 \\
    \bottomrule
\end{tabular}
\caption{Impact of the choice of loss.}
\label{tab:loss}

\end{minipage}
\hfill
\begin{minipage}[t]{0.32\textwidth}
\centering
% \begin{table}[t]
\centering
\small{
\begin{tabular}{c ccc}
    \toprule
    $K_\text{max}$ & mAP & HOTA & FSLA \\
    \midrule
    2              & 46.68 & 65.12 & 46.48 \\
    3              & 45.23 & 64.67 & 44.90 \\
    4              & 44.94 & 64.01 & 44.06 \\
    \bottomrule
\end{tabular}}
\vspace{-0.6em}
\caption{Impact of $K_\text{max}$.}
\label{tab:kmax}
\end{minipage}
\label{tab:all_results}
\vspace{-0.9em}
\end{table*}

\vspace{-0.4em}
\subsection{Experimental Settings}
\vspace{-0.1em}
\paragrapht{Datasets.}
We evaluate on AVISeg benchmark containing 926 videos (16 hours, 61.4s average), 94,074 instance masks across 56,871 frames in 26 categories.
Each video is divided into 1 fps clips, and only objects that emit sound are exhaustively annotated with persistent identifiers while silent instances are not masked.

\paragrapht{Metrics.}
Following~\cite{guo2025avis}, we report three primary metrics that jointly assess detection, localization, and identity association over time.
While mAP~\cite{yang2019vis} is computed on video trajectories using spatio–temporal IoU, HOTA~\cite{luiten2021hota} jointly measures detection and association through frame-wise bijective matching and sequence-level scoring.
FSLA~\cite{guo2025avis} is the fraction of correct frames after bipartite matching, requiring matched counts/categories and per-object IoU $\ge \alpha$, averaged over $\alpha \in \{0.05,0.10,\dots,0.95\}$.
We also report FSLA for silent, single-source, and multi-source frames (FSLAn/s/m).

\paragrapht{Implementation details.}
Following AVISM protocol, we use ResNet-50~\cite{he2016resnet} for visual and VGGish~\cite{hershey2017vggish} for audio.
% Video frames are sampled at 1 FPS and audio at 16 kHz.
We resize the shorter image side to 360 pixels during training and 448 pixels at inference, keeping aspect ratio.
Our model uses $N_f=100$ frame queries and $N_v=100$ video queries with a window size $W=6$ in the video-level tracker.
We set all loss weights to 1.0, except the weight of $\mathcal{L}_{\text{sim}}$, which is 0.5.
% The experiments are conducted on 2 NVIDIA A5000 GPUs.
\vspace{-1.1em}
\subsection{Main Results}
\vspace{-0.5em}
As shown in Table~\ref{tab:main}, our \method improves overall detection and tracking quality over the baseline, raising mAP from 45.04 to 46.68 (+1.64), HOTA from 64.52 to 65.12 (+0.60), and FSLA from 44.42 to 46.48 (+2.06).
On the decomposed FSLA scores, our method favors sounding-object localization in both single-source and multi-source frames, with gains on FSLAs (+1.83) and FSLAm (+3.82).
Figure~\ref{fig:qualitative} shows reduced mask coalescence and identity swaps in crowded scenes.

\vspace{-1.1em}
\subsection{Ablation Studies}
\vspace{-0.5em}
\paragrapht{Audio-centric query generator and $\mathcal{L}_{\text{SAOC}}$.}
Table~\ref{tab:acqg} shows that adding $\mathcal{L}_{\text{SAOC}}$ to ACQG improves all metrics: mAP 45.17 to 46.68 (+1.51), HOTA 63.27 to 65.12 (+1.85), and FSLA 45.45 to 46.48 (+1.03).
This indicates that audio-conditioned queries and ordinal counting provide complementary benefits.

\paragrapht{Loss design.}
Replacing standard cross-entropy with $\mathcal{L}_{\text{SAOC}}$ consistently improves performance (Table~\ref{tab:loss}): mAP 44.45 to 46.68 (+2.23), HOTA 63.95 to 65.12 (+1.17), and FSLA 44.00 to 46.48 (+2.48).
The ordinal formulation better ranks hypotheses, stabilizing matching and reducing identity switches.

\paragrapht{Backbone and pretraining dataset.}
Pretraining on ImageNet~\cite{deng2009imagenet} and COCO~\cite{lin2014coco}, rather than ImageNet alone, yields clear improvements (Table~\ref{tab:backbone}): mAP 42.14 to 46.68 (+4.54), HOTA 62.09 to 65.12 (+3.03), and FSLA 42.87 to 46.48 (+3.61).
Replacing ResNet-50 with Swin-L~\cite{liu2021swin} is expected to further improve segmentation quality and long-range identity maintenance.

% \paragrapht{Number of queries.}
% Increasing queries from 25 to 100 significantly improves all metrics (Table~\ref{tab:query}): FSLA +6.93, HOTA +2.59, and mAP +2.56.
% More queries better disentangle concurrent sources and cover transient sounders.

\paragrapht{Sensitivity to $K_\text{max}$.}
Table~\ref{tab:kmax} analyzes different $K_{\max}$ values for SAOC loss. Setting $K_{\max}=2$ performs best, aligning with the dataset's typical sounding object distribution, while higher values reduce accuracy.

\label{sec:pagestyle}

\vspace{-0.7em}
\section{Conclusion}
\vspace{-0.5em}
\label{sec:typestyle}
% We presented audio-centric query generation and sound-aware ordinal counting loss to address visual dominant regime in audiovisual instance segmentation.
% Our method enables queries to specialize to different sound sources, while SAOC loss explicitly supervises sounding object counts via ordinal regression.
% Comprehensive experiments demonstrate that both query specialization and  counting supervision are essential, achieving strongest gains in challenging multi-source scenario.
% These results validate that balanced frame-level processing via both architectural and objective design is crucial for accurate audiovisual instance segmentation.

% { \color{blue}
% We propose audio-centric query generation and a sound-aware ordinal counting loss to mitigate visual dominance in audiovisual instance segmentation.
% Our approach specializes queries for different sound sources and supervises sounding-object counts via ordinal regression, yielding strong gains in challenging multi-source scenario.

% We propose audio-centric queries and sound-aware ordinal counting loss to address visual dominance in audiovisual instance segmentation.
We propose audio-centric queries and sound-aware ordinal counting loss to address visual dominance in AVIS.
By specializing queries and supervising counts, our method achieves significant gains in multi-source scenarios, validating the importance of balanced frame-level processing.

% \vspace{-0.7em}
% { \color{blue}
\paragrapht{Acknowledgement.}
% \vspace{-0.5em}
% This work was supported by the National Research Foundation of Korea(NRF) grant funded by the Korea government(MSIT) (RS-2025-00515741).
This work was supported by the National Research Foundation of Korea(NRF) grant funded by the Korea government(MSIT) (RS-2025-02216328).
% }

\vfill\pagebreak


\bibliographystyle{IEEEbib}
\small{\bibliography{strings}}

\end{document}